\documentclass[aip, amsmath, amssymb,reprint,pop,floatfix]{revtex4-2}
\usepackage{amsfonts}
\usepackage[utf8]{inputenc}
\usepackage{graphicx} 

\usepackage{comment}
\usepackage[hidelinks]{hyperref}
\usepackage{xcolor}
\newcommand{\pdv}[2]{\frac{\partial #1}{\partial #2}}
\newcommand{\dv}[2]{\frac{\mathrm{d} #1}{\mathrm{d} #2}}

\newcommand{\changesI}[1]{\textcolor{blue}{#1}}

\begin{document}

\title{Flowing plasma rearrangement in the presence of static perturbing fields} 

\author{T. Rubin}
\email{trubin@princeton.edu}
\affiliation{Department of Astrophysical Sciences, Princeton University, Princeton, New Jersey 08540, USA}

\author{I. E. Ochs}
\affiliation{Department of Astrophysical Sciences, Princeton University, Princeton, New Jersey 08540, USA}

\author{N. J. Fisch}
\affiliation{Department of Astrophysical Sciences, Princeton University, Princeton, New Jersey 08540, USA}
\date{\today}
\begin{abstract}
	Charged particles interacting with electromagnetic waves have a portion of their energy tied up in wave-driven oscillations. When these waves are localized to the exhaust of linear magnetic confinement systems this ponderomotive effect can be utilized to enhance particle confinement. The same effect can be derived for particles moving via an $\mathbf{E} \times \mathbf{B}$ drift into a region of a static perturbation to the electromagnetic fields which has a large wave vector component in the direction of the motion. In this work we use a simplified slab model to self-consistently solve for the electromagnetic fields within the fluid flowing plasma of a static flute-like ($k_\parallel = 0)$ perturbation, and evaluate the resulting ponderomotive potential. We find that two types of perturbations can exist within the flowing plasma, which are an O wave and an X wave in the frame moving with the fluid. In the case of tenuous plasma, these perturbations are magnetostatic or electrostatic multipole-analog perpendicular to the guiding magnetic field in the lab frame, respectfully. For denser plasmas, the O wave-like perturbation is screened at the electron skin depth scale, and the X wave-like perturbation is a combination of a similar perpendicular electric perturbation and parallel magnetic perturbation. The ponderomotive potential generated in the X wave-like case is gyrofrequency-dependent, and can be used as either potential barriers or potential wells, depending on the direction of the flow velocity.
\end{abstract}

\maketitle
\section{Introduction}

{Interest in magnetic mirror machines for nuclear fusion applications is increasing as several groups are attempting to realize modern magnetic mirrors\cite{burdakovMultiplemirrorTrapMilestones2016,fowlerNewSimplerWay2017,beeryPlasmaConfinementMoving2018a,whiteCentrifugalParticleConfinement2018,ivanovGasdynamicTrapOverview2013a,endrizziPhysicsBasisWisconsin2023,schwartzMCTrans0DModel2023}. }
{Plasma rotation in a linear magnetic confinement device can provide additional axial confinement. This is done by employing concave drift-surfaces that rotate at a near-uniform rotation frequency and the constancy of the canonical angular momentum to repel particles from regions of large-curvature, in the same manner a bead is pulled to the center of a rotating string.
 Generating rotation of this kind, however, necessitates an investment of the kinetic energy associated with the rotation for each particle. This energy cost scales with the size of the required energy barrier. When seeking to confine high-energy plasma, such as a proton-boron 11 fusion plasma, other confining potentials may be required.}

In addition, the plasma may be plugged by RF ponderomotive quasipotential\cite{gormezanoReductionLossesOpenended1979,postMagneticMirrorApproach1987,millerRFPluggingMultimirror2023,dimontePonderomotivePseudopotentialGyroresonance1982}, which is the secular result of an oscillating field on the particle dynamics\cite{gaponov1958potential, motzRadioFrequencyConfinementAcceleration1967,lichtenbergRegularStochasticMotion1983}. The term ``quasipotential" is used here in order to stress the point that this effect is an emergent property of the particle dynamics and is only applicable under specific circumstances. In addition, like the diamagnetic ``mirror" potential, the ponderomotive effect applies differently to particles with different phase-space coordinates, such as different gyro radius. We will use the term ``potential" rather than ``quasipotential" for brevity.

Oscillating fields can generate a ponderomotive potential, if there exists a separation of time scales between the field oscillation and the field envelope variation at the particle position, and if the particle is not resonant with the field. A magnetized charged particle gyrates around a magnetic field line. This gyration degree of freedom has a natural frequency - the cyclotron frequency. As such, the expression for the leading order ponderomotive potential for a field interacting with this degree of freedom would depend on the cyclotron frequency and the squared amplitude of the field component.
 
The ponderomotive effect can be used to repel or attract\cite{dodinApproximateIntegralsRadiofrequencydriven2005,dodinNonadiabaticPonderomotivePotentials2006,dodinPonderomotiveBarrierMaxwell2004} particles from regions of positive or negative potential, hold them in place or manipulate them in some more complex way\cite{chuNobelLectureManipulation1998, cohen-tannoudjiNobelLectureManipulating1998}. {These ponderomotive potentials can, in principle, also be achieved in rotating or flowing plasma through static perturbations\cite{rubinGuidingCentreMotion2023a, rubinMagnetostaticPonderomotivePotential2023a,ochsCriticalRoleIsopotential2023b}, which would be seen as waves in the rotating frame. The advantage of static perturbations is that RF power need not be used to generate the ponderomotive potential.} 

Due to the charge and mass dependence of these ponderomotive barriers, they are also used to selectively confine one ion species over another\cite{hidekumaPreferentialRadiofrequencyPlugging1974,hiroeRadiofrequencyPreferentialPlugging1975}, which is of use in the related field of isotope separation\cite{weibelSeparationIsotopes1980}.

{Setting up the plasma rotation in a magnetic mirror machine is often done using concentric end-electrodes\cite{bekhtenevProblemsThermonuclearReactor1980}, each of which is biased to a different electric potential. The potential at the end-electrodes is conducted into the center of the device by the large plasma conductivity along drift surfaces. Different drift surfaces remain at different potentials due to the much smaller plasma conductivity across them.
Plasma drift surfaces follow closely the geometry of magnetic surfaces. Thus, magnetic surfaces tend to have near-constant electric potential. This is the initial assumption used in the isorotation theorem\cite{ferraroNonuniformRotationSun1937} to generate uniform rotation frequency along a flux surface. This argument is used in the ideal magnetohydrodymanics approximation to set the electric field parallel to the magnetic field to zero.}

Several recent works \cite{rubinGuidingCentreMotion2023a, rubinMagnetostaticPonderomotivePotential2023a,ochsCriticalRoleIsopotential2023b} have investigated the behavior of a flowing magnetized plasma interacting with a static magnetic perturbation. In Refs.~\onlinecite{rubinGuidingCentreMotion2023a, rubinMagnetostaticPonderomotivePotential2023a}, which considered realistic magnetic fields inside a cylinder, including the boundary conditions used to generate them, where the plasma was assumed to be tenuous enough not to affect fields the ponderomotive potential was found to be positive definite, since the $v\mathbf{e}_\theta\times\mathbf{B}_1$ force in this configuration doesn't interact with the gyromotion to leading order.
In Ref.~\onlinecite{ochsCriticalRoleIsopotential2023b}, a simplified slab geometry was used to consider two different {electric and magnetic field configurations: (i) a linearly-polarized magnetic perturbation, perpendicular to $\mathbf{B}_0$}, where the electric field was not affected by the magnetic perturbation, and (ii) an MHD-like limit, where electrons were assumed to short the potential along magnetic field lines, {and the electric potential remained a constant on each magnetic field line}. 
The ponderomotive potential was shown to be very different in each case.
Taken together, the three studies demonstrated the necessity of solving for the electromagnetic fields self-consistently in the presence of the flowing plasma, in order to determine the ponderomotive potentials which can be employed in a practical device.

Thus, in this work, we consistently solve for the wave propagation of a flute-like ($k_\parallel = 0$) mode into an $\mathbf{E} \times \mathbf{B}$ flowing plasma.
As in Ref.~\onlinecite{ochsCriticalRoleIsopotential2023b}, we work in the slab limit, which allows the lab-frame boundary conditions to be boosted into the flowing frame, where the fluid equations are very simple to solve{, and the perturbation indeed becomes a time-dependent wave}.
We are thus able to solve self-consistently for the wave penetration and ponderomotive potential.

{We find that in the moving frame, the two branches of the dispersion relation are self-consistent solutions to the plasma response, which are O-waves and the X-waves. The O-wave in the moving frame corresponds to a perpendicular multipole-analog magnetic field in the lab, and is the perturbation investigated in Refs.~\onlinecite{rubinGuidingCentreMotion2023a, rubinMagnetostaticPonderomotivePotential2023a}. We find that in a flowing plasma, the magnetic perturbations are cut off at the electron skin depth, which tends to be short in all but the most tenuous plasmas, thus limiting their application to isotope separators and other extreme-vacuum laboratory devices. In contrast, the X-wave in the moving frame has enhanced penetration} into the flowing plasma, and may even propagate rather than be evanescent. 
These perturbations do couple to the internal degree of freedom, and generate flow-dependent positive or negative ponderomotive potential, as found in \citet{dodinApproximateIntegralsRadiofrequencydriven2005,ochsCriticalRoleIsopotential2023b}. {
As a result, the same perturbation can generate a ponderomotive effect with opposite sign for electrons and ions.}

This paper is organized as follows; In Section~\ref{sec:2} we describe the cold plasma waves propagating perpendicularly to the magnetic field. In Section~\ref{sec:3} we utilize Lorentz boost to derive the electromagnetic fields the flowing plasma can support, and the boundary conditions for these perturbations in the lab frame. In Section~\ref{sec:4} we derive the kinetic ponderomotive potentials which appear in the guiding center frame, and discuss their attractive or repulsive effect on particles. In Section~\ref{sec:5} we present numerical results validating the calculations in the previous sections using a full-orbit code. 

{\section{Cold plasma waves}\label{sec:2}
In this section, we look at the wave modes existing in a magnetized cold uniform stationary (i.e. with zero flow velocity) fluid plasma, where their time-dependence could be eliminated by boosting to a frame moving with a velocity $-v\mathbf{e}_{y'}$. In the frame where these waves become time independent, the plasma flows with the velocity $v\mathbf{e}_{y}$, and the Lorentz-boosted zero-frequency ``waves" describe the static perturbations that can exist in the linear regime in the flowing plasma. We denote the frame in which the waves are time-dependent with a prime (the moving frame), and the frame in which the perturbation is time-independent without a prime (the lab frame).}

{The simplest plasma we can consider is a uniform cold fluid plasma. The response of this plasma to small-amplitude linear electromagnetic waves is described by the dispersion relation\cite{stixWavesPlasmas1992}
\begin{gather}
    \begin{pmatrix}
        S-N_{y'}^2-N_{z'}^2 & -iD+ N_{x'}N_{y'} & N_{x'}N_{z'}\\
        iD+N_{x'}N_{y'} & S-N_{x'}^2-N_{z'}^2 & N_{y'}N_{z'}\\
        N_{x'}N_{z'}& N_{y'} N_{z'}& P-N_{x'}^2 -N_{y'}^2
    \end{pmatrix} \mathbf{h} '
    =0,
\end{gather}
with
\begin{gather}
    S = \frac{1}{2}(R+L), \ \ 
    D = \frac{1}{2}(R-L),\\
    R,L=1-\sum_s \frac{\omega_{ps}'^2}{\omega'(\omega'\pm\omega_{cs}')}, \ \ P = 1- \sum_s \frac{\omega_{ps}'^2}{\omega'^2}.
\end{gather} 
Where $\mathbf{N}'=\mathbf{k}'c/\omega'$ is the refractive index for a wave with a wave vector $\mathbf{k}'$ and a frequency $\omega'$ in the primed frame, $\mathbf{h}'$ is the electric field (complex) polarization vector, $\omega_{ps}'^2 = Z_s^2 e^2 n_{s'}/\epsilon_0 m_s$ is the plasma frequency of species $s$, $n_{s}'$ is its number density in the primed frame, and $\omega_{cs}' = Z_s e B_0' / m_s$ is its cyclotron frequency in the primed frame.}

{We consider the ``slab analog" of the system described in \citet{rubinGuidingCentreMotion2023a, rubinMagnetostaticPonderomotivePotential2023a}. This is achieved by taking the $x$ direction to be analogous to the radial ($r$) direction, and the $y$ direction to be analogous to the azimuthal ($\theta$) direction in a cylinder, with the $z$ direction remaining the same. The uniform rotation around the axis of the cylinder is replaced with a uniform flow along the $y$ coordinate, and we examine the half volume $x<0$ in which the plasma resides. A static magnetic field $B_0'\mathbf{e}_z$ is the analog to the axial field in a mirror machine.}

In order to eliminate the wave time-dependence using a frame transformation to a frame moving in the $y'$ direction, we must take $N_{y'}\neq 0$. In order to generate a ponderomotive potential in the $z'$ direction, we are interested in waves with a slowly varying envelope in the $z'$ direction, starting from zero outside the perturbation. There is no constraint on the value of $k_{z'}$, and the value of $k_{x'}$ is determined by the plasma response, using the dispersion relation when the perturbation is small enough for the dispersion relation to be valid.

The case where $k_{z' }= N_{z'} =0$ is algebraically simplest, because the dispersion relation can be decomposed into the O wave and the X wave.

The dispersion relation for the O wave is 
\begin{gather}
	P -N_{x'}^2-N_{y'}^2=0,\label{eq:O disp}
\end{gather}
with an electric field polarization
\begin{gather}
	\mathbf{h} = \mathbf{e}_{z'}.
\end{gather}

{The dispersion relation for the X wave is 
\begin{gather}
	S^2-(N_{y'}^2+N_{x'}^2)S- D^2=0,\label{eq:X disp}
\end{gather}
with an electric field polarization 
\begin{gather}
	\frac{h_{x'}}{h_{y'}} = \frac{iD-N_{x'}N_{y'}}{S-N_{y'}^2},\ \ h_{z'}  =0. 
\end{gather}}

{All perturbations with $k_{z'}=0$, $N_{y'}\neq0$ must be some linear composition of O waves and X waves in the moving frame. The value of $N_{y'}$ is dictated by the boundary condition on the $x=0$ plane and the velocity $v$, and the value of $N_{x'}$ is determined from the dispersion relation.}

\section{Transformation from the moving frame to the lab frame}\label{sec:3}
If  $|v|<c$, we can use the Lorentz transformation, in order to move to the frame moving with the flow.  
The Lorentz boost from a Cartesian lab frame coordinates $x^\mu=(ct,x,y,z)$ to a moving frame with coordinates $x^{\mu'} = (ct',x',y',z')$, traveling in the $y$ direction with a velocity $v = \beta c$, in a flat spacetime with a metric $\eta =\mathrm{diag}(-,+,+,+)$, and a Lorentz factor $\gamma = (1-\beta^2)^{-1/2}$ is
\begin{gather}
    x^{\mu'} = \Lambda^{\mu'}_{\ \nu} x^\nu,\\
    \Lambda^{\mu'}_{\ \nu} =
    \begin{pmatrix}
        \gamma& 0& -\gamma \beta& 0\\
        0& 1 & 0&0\\
        -\gamma \beta & 0 &\gamma&0\\
        0& 0& 0& 1
    \end{pmatrix},
\end{gather}
\begin{gather}
    x^{\mu'} =
    \begin{pmatrix}
        \gamma(ct-\beta y)\\
        x\\
        \gamma(y-vt)\\
        z
    \end{pmatrix}, \ \ x^{\mu} = 
    \begin{pmatrix}
        \gamma(ct'+\beta y')\\
        x'\\
        \gamma(y'+vt')\\
        z'
    \end{pmatrix}.
\end{gather}

The inverse transformation $\Lambda^{\nu}_{\ \mu'}$ is the Lorentz transformation to a frame moving with the opposite velocity: $-v = -\beta c$ so $x^\alpha = \Lambda^{\alpha}_{\ \mu'} x^{\mu'}=\Lambda^{\alpha}_{\ \mu'}\Lambda^{\mu'}_{\ \nu} x^\nu$.

The wave 4-covector
\begin{gather}
	k_\mu = \begin{pmatrix}
		-\omega/c &\mathbf{k}
	\end{pmatrix},
\end{gather}
transforms as a covector
\begin{gather}
    k_{\mu'} = k_\nu \Lambda^{\nu}_{\ \mu'} = \begin{pmatrix}
		\gamma(v k_y-\omega)/c &k_x & \gamma (k_y-\beta \omega /c) & k_z
	\end{pmatrix},\label{eq:k trans}
\end{gather}
in order to retain the invariace of the phase. For a perturbation with $\omega=0$ and $k_y\neq0$, the wave vector components in the moving frame are $k_{x'}=k_x$, $k_{y'} = \gamma k_y$, $\omega' = -\gamma k_y v$.

The 4-potential field in the moving frame is
\begin{gather}
    A_{\mu'} = \begin{pmatrix}
        -\Phi'/c& \mathbf{A}'
    \end{pmatrix},
\end{gather}
and in the lab frame,
\begin{gather}
    A_{\mu}  = \begin{pmatrix}
		-\gamma(\Phi'+v A_{y'})/c &A_{x'} & \gamma (A_{y'}+\beta \Phi' /c) & A_{z'}
	\end{pmatrix}.
\end{gather}

{Using these transformations, we confirm that the vector potential 
\begin{gather}
	A_{\mu'} = \begin{pmatrix}
		0 & 0& x' B_0'& 0
	\end{pmatrix},
\end{gather}
which generates a $\mathbf{B}' = B_0'\mathbf{e}_{z'}$ in the moving frame, is expressed in the lab as
\begin{multline}
    A_{\mu}  = \begin{pmatrix}
		-\gamma(v x' B_0')/c &0 & \gamma x' B_0' &0
	\end{pmatrix}\\ = \begin{pmatrix}
		x E_0/c &0 &x B_0 &0
	\end{pmatrix},
\end{multline}
for $B_0' = B_0/\gamma$ and $v = -E_0/B_0$.}

For practical reactor applications, it is unlikely to exceed plasma flow velocity much in excess of $\pm0.1 c$.

\subsection{O Wave Dispersion}

An O wave with with $k_{y'} = \gamma k$ and $\omega' = - k \gamma v$ has an $N_{y'} =- \gamma k c/ k \gamma v = -1/\beta$ has to satisfy the dispersion relation in equation (\ref{eq:O disp})
\begin{gather}
    N_{x'}^2=P-\beta^{-2} ,\\
    k_{x'}^2= -i \sqrt{k^2+\sum_s \frac{\omega_{ps}'^2}{c^2}}=-ik\kappa_O,\\
    \kappa_O = \sqrt{1+\sum_s \frac{\omega_{ps}'^2}{c^2k^2}}.\label{eq:k o}
\end{gather}
where we selected the solution that is evanescent for negative $x'$, corresponding to the boundary conditions on the $x'=0$ plane. We find that the O wave spatial decay is enhanced by the introduction of the skin-depth length-scale, $c/\omega_{pe}$.

The O wave in the moving frame has the following electromagnetic fields,
\begin{gather}
    \mathbf{E}'_O = \Re\left[ -iE_1'e^{k\kappa_O x'+ik\gamma (y'+vt')}\right]\mathbf{e}_{z'},\label{eq:E1O}\\
    \mathbf{B}'_O = \Re\left[\frac{E_1'}{\gamma v}e^{k \kappa_O x'+ik \gamma (y'+vt')}\left(i\gamma\mathbf{e}_{x'}-\kappa_O\mathbf{e}_{y'}\right)\right],
\end{gather}
with $E_1'$ being a (complex) amplitude which may be a function of $z$, and $\Re[f]$ denotes the real part of $f$. This mode is characterized by a perpendicular propagation, $\mathbf{k}\cdot\mathbf{B}_0=0$, with $\mathbf{E}\cdot\mathbf{B}\neq0$ - in the moving frame, the electric field component of the wave is polarized along $\mathbf{B}_0$. The quantity $\mathbf{E}\cdot\mathbf{B}$ is a Lorentz invariant, which is independent of the frame in which the electromagnetic fields are observed.

Selecting a gauge, this time-dependent wave can be generated by the following vector potential
\begin{gather}
	\mathbf{A}'_O = -\int \mathbf{E}'_O dt' = \Re\left[\frac{E_1'}{ k \gamma v}e^{k\kappa_O x'+i k \gamma (y'+vt')}\right]\mathbf{e}_{z'},
\end{gather}
and a $\Phi'_O=0$.

Looking at this 4-potential in the lab frame, and transforming the coordinates $x'=x$ and $y=\gamma (y'+vt')$,
\begin{gather}
    A_{O\mu}  = \Re\left[\frac{E_1'}{\gamma k v}\begin{pmatrix}
		0 &0 &0 &1
	\end{pmatrix}e^{k\kappa_O x'+i ky}\right].
\end{gather}
This is a vector potential generating a purely magnetic field, which is
\begin{gather}
	\mathbf{B}_O = B_1 e^{k\kappa_O x}\left(\sin ky \mathbf{e}_x+\kappa_O \cos ky \mathbf{e}_y\right),\label{eq:b1va}
\end{gather}	
with $B_1 = E_1'B_0/ \gamma E_0$. In the lab frame, the electric field component of the perturbation disappears, so one can't define the electric field polarization in the lab frame. The magnetic field polarization in the lab frame is elliptical, and dominated by the $\mathbf{e}_y$ component.

This is the correct Cartesian-coordinate analog to a magnetic multipole. The boundary conditions in the lab frame perturbation are indeed purely magnetic and are generated with current-carrying coils in the lab along the $z$ direction.

In a tenuous plasma, where $n_s\approx0$, this is the correct analog to the magnetic multipole perturbation used in \citet{rubinGuidingCentreMotion2023a, rubinMagnetostaticPonderomotivePotential2023a}. For a typical fusion reactor, we expect an electron density of $n_e=10^{14} [cm^{-3}]$. Taking $k = n/R = 2 [m^{-1}]$, substituting in equation (\ref{eq:k o}) to have the value $k = -i \sqrt{k^2+ e^2 n_e/\epsilon_0 \gamma m_e c^2} = -i\sqrt{2\cdot 10^{-4}+ 10^{14} / (5.31^2\cdot 10^{10})} [cm^{-1}]= - 18.8[cm^{-1}] i$. This is a fast decay rate, for which the main contribution is the skin depth.

The conclusions from this part of the calculation are \textit{i}) the plasma rejects this magnetic perturbation, rather than enhancing it. This rejection can be quite significant for denser (fusion) plasma. These perturbations can never propagate in a fluid plasma no matter its density. \textit{ii}) When applying a pure magnetic perturbation that is perpendicular to the guide field to a flowing plasma, the plasma does not rearranges itself such as to modify the electric fields, but modifies the magnetic fields instead. {The viability of linearly polarized magnetic perturbations is discussed in Appendix~\ref{app:1}.}

\subsection{X Wave Dispersion}
An X wave with the same $k_{y'} = \gamma k$, $\omega' = - k \gamma v$ and $N_{y'} =- \gamma k c/ k \gamma v = -1/\beta$ has to satisfy the dispersion relation in equation (\ref{eq:X disp}). This mode is characterized by a perpendicular propagation, $\mathbf{k}\cdot\mathbf{B}_0=0$, with $\mathbf{E}\cdot\mathbf{B}=0$. 

When
\begin{gather}%
	N_{x'}^2=S-\frac{D^2}{S}-\frac{1}{\beta^2}<0,
\end{gather} 
the X wave is evanescent in $x'$, and for $x<0$
\begin{gather}
	k_{x'} =-ik\sqrt{\beta^2\gamma^2 \left|N_{x'}^2\right|}=-ik \kappa_X,\\
	\kappa_X = \sqrt{1-\beta^2\gamma^2 (S-1)+ \beta^2\gamma^2\frac{D^2}{S}}\label{eq:kappa X},
\end{gather}
Even for a non-propagating wave, the decay length is generally much longer than for the O wave.
The wave polarization is elliptical in the ${x'}-{y'}$ plane is
\begin{gather}
	p=\frac{h_{x'}}{h_{y'}} = i\frac{D+ \frac{\kappa_X}{\gamma\beta^2}}{S-\frac{1}{\beta^2}},\ \ h_{z'}  =0. \label{eq:pol kappa x}
\end{gather}

When $N_{x'}^2>0$ the X wave can propagate in the plasma. Considering waves propagating in the negative $x$ direction, (with $N_{x'}<0$), yields the $x'$ wave vector component
\begin{gather}
	k_{x'} =k  \sqrt{\beta^2\gamma^2 N_{x'}^2}= k \tilde{k}_X.\label{eq:tilde k X}
\end{gather}
The polarization of a propagating X wave is 
\begin{gather}
	p=\frac{h_{x'}}{h_{y'}} = \frac{iD- \frac{\tilde{k}_X }{\gamma \beta^2}}{S-\frac{1}{\beta^2}},\ \ h_{z'}  =0. \label{eq:pol k x}
\end{gather}
The expressions for $\kappa_X$ and $\tilde{k}_X$ are even in $v$ (or $\beta$).

\subsubsection{The Low-Frequency Limit}
The expressions for $S$ and $D$ are greatly simplified in the low-frequency limit, where $|\omega'|=|\gamma kv|\ll \omega_{ci}'$, for the smallest cyclotron frequency among species in the plasma. In this limit,
\begin{gather}
	S = 1+\gamma_A,\ \ D=0,\\
	\kappa_X = \sqrt{1-\beta^2\gamma^2\gamma_A},\label{eq:kappa X LF}\\
   	p_{LF}=\frac{h_{x'}}{h_{y'}} = \frac{\gamma }{i\kappa_X}\label{eq:p LF}
\end{gather}
with $\gamma_A=\sum_s\omega_{ps}'^2/\omega_{cs}'^2$. This mode is evanescent if $\gamma^2 \beta^2 \gamma_A < 1$. This approximation if well suited for use in case of low $k$.
The cutoff in $N_{x'}$ appears when $\beta^2\gamma^2\gamma_A=1$, beyond which the wave can become propagating in the moving frame.

In the nonrelativistic limit, where $\gamma\approx 1$,
\begin{gather}
	\beta^2 \gamma_A = \sum_s\frac{v^2m_s n_s\mu_0}{B_0^2} =\frac{v^2}{  v_A^2} = M_A^2,
\end{gather}
this term is the Alfv\'en Mach number squared. Thus, as found for MHD waves\cite{fettermanAlphaChannelingRotating2010b}, low-frequency X waves can only propagate if the plasma is rotating supersonically with respect to the Alfv\'en speed. 

In general, the wave electric field with a polarization $h_{x'}/h_{y'}= p\in\mathbb{C}$ is 
\begin{gather}
	\mathbf{E}'_{X} = \Re\left[iE_1\frac{p\mathbf{e}_{x'}+\mathbf{e}_{y'}}{\sqrt{|p|^2+1}}e^{k \kappa_X x'+i k\gamma(y' +vt')}\right].
\end{gather}
The wave electric field and its polarization coefficients in the moving frame can be written as
\begin{multline}
	\mathbf{E}'_{X} =  \Re\Big[-E_1\left(E_{L'}\mathbf{e}_{L'}+E_{R'} \mathbf{e}_{R'}-iE_{x'}\mathbf{e}_{x'}\right)\\\times e^{k \kappa_X x'+i k\gamma(y' +vt')}\Big],
\end{multline}
\begin{gather}\nonumber
	E_{L'}=\frac{1}{\sqrt{2}}\frac{\Im[p]-1}{\sqrt{|p|^2+1}}, \ \ 	E_{R'}=\frac{1}{\sqrt{2}}\frac{1+\Im[p]}{\sqrt{|p|^2+1}},\\ E_{x'} = \frac{\Re[p]}{\sqrt{|p|^2+1}},\label{eq:polare}
\end{gather}
corresponding to the left and right polarization unit vectors as $\mathbf{e}_{R'} = (\mathbf{e}_{x'}-i\mathbf{e}_{y'})/\sqrt{2}$, $\mathbf{e}_{L'} = (\mathbf{e}_{x'}+i\mathbf{e}_{y'})/\sqrt{2}$. Here, $\Im[f]$ denotes the imaginary part of $f$. With $p$ defined in equations (\ref{eq:pol kappa x}) and (\ref{eq:pol k x}) for the evanescent and propagating regimes, respectfully.

\begin{figure}
	\centering
	\includegraphics[width = \columnwidth]{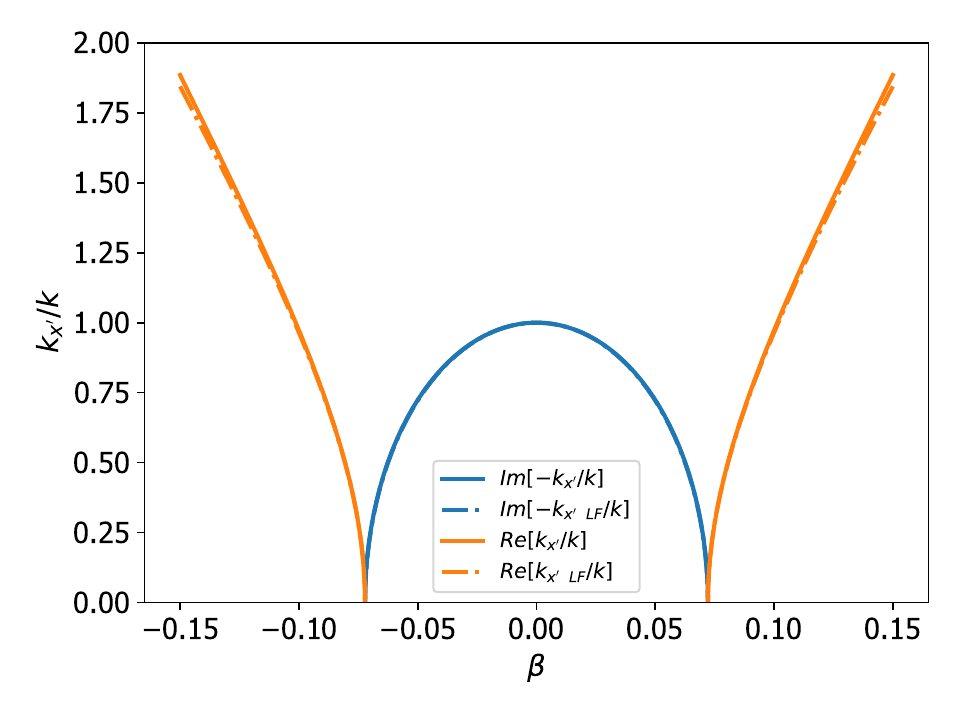}
	\caption{The exact dispersion relation, $k_{x'}$ as a function of $\beta = v/c$, for a quasi-neutral electron-proton plasma, with $n_e =n_p = 10^{20}[m^{-3}]$, $B_0 = 10[T]$, and $k = 50[m^{-1}]$, plotted in solid lines. The cutoffs appears in $\beta =\pm0.07$, and between them the wave is evanescent. Away from the cutoffs, the wave is propagating. The low-frequency approximation is plotted in dashed lines, showing excellent fit in the entire plotted region.}\label{fig:k50ep plasma}
\end{figure}

\begin{figure}
	\includegraphics[width = \columnwidth]{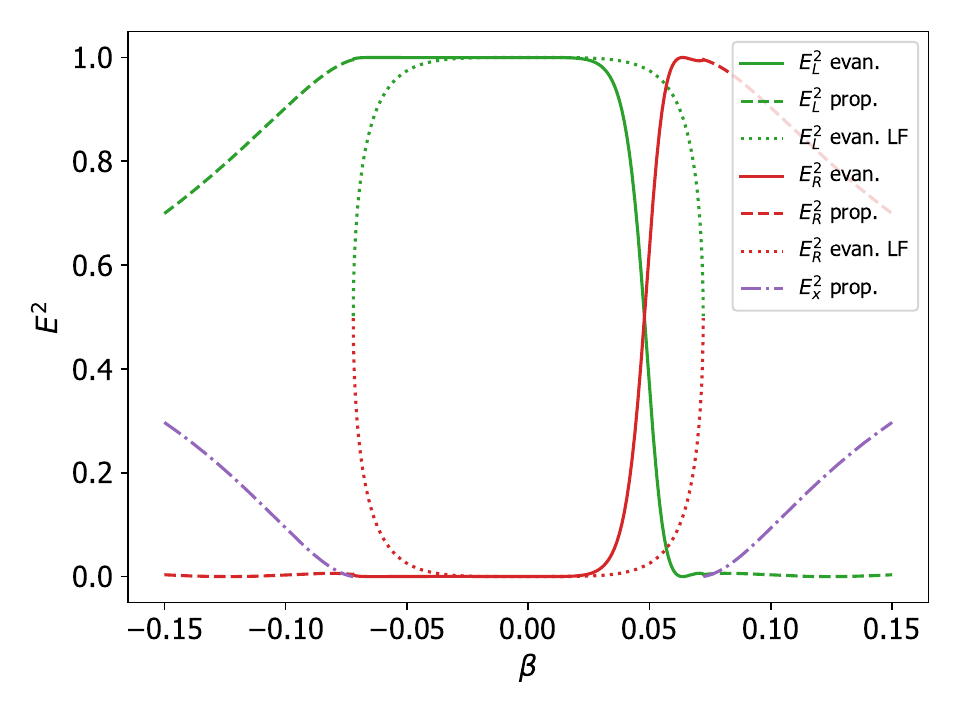}
	\caption{Wave polarization squared in the moving frame, as a function of $\beta$, for the same parameters defined in Figure \ref{fig:k50ep plasma}. Plotted are the squares of the coefficients of the unit vector in the electric field direction. In the evanescent regime, the wave is composed of only right and left polarization, without a phase shift. In the propagating regime, the wave acquires a phase-shift, essentially splitting into three components.} 
\label{fig:k50ep plasma pol}
\end{figure}

In Figure \ref{fig:k50ep plasma} we plot the dispersion relation for an electron-proton quasineutral plasma, with different curves representing the exact expressions for the evanescent and propagating regimes from equations (\ref{eq:kappa X}) and (\ref{eq:tilde k X}), as well as the approximation in equation (\ref{eq:kappa X LF}). The wave in the moving frame is in the low frequency limit when $\beta\ll 0.064$ for the parameters chosen for this plot. However, the expression for $k_{x'}$ remains fairly accurate for a significantly larger range of $\beta$. In Figure \ref{fig:k50ep plasma pol} we plot the squares of the coefficients of the wave, for the left, right and linear polarizations. In the low frequency limit we replace $p$ with $p_{LF}$ from equation (\ref{eq:p LF}) in equations (\ref{eq:polare}). In this figure, we see that the polarization deviates significantly from the low frequency approximation around $|\omega'|\approx \omega_{ci}$.


In the low frequency limit, the wave electric field is
\begin{gather}
    \mathbf{E}'_{\mathrm{X}} \propto\Re\left[\gamma\mathbf{e}_{x'} + i\kappa_X\mathbf{e}_{y'}e^{k \kappa_X x'+i k\gamma(y' +vt')}\right].
\end{gather}
This is an elliptically (mostly left) polarized wave in the moving frame, corresponding to a linear combination of two circularly polarized waves with the two polarizations. A pure circular polarization is impossible in the low frequency limit, except in the trivial $v=0$ case, as it requires $\kappa_X=\gamma$, but $\kappa_X\leq1$ and $\gamma\geq1$ for all $v\neq0$.  

A vector potential generating the perturbation is $\mathbf{A}'_X = -\int \mathbf{E}'_{\mathrm{X}} dt'$, or in a 4-covector form
\begin{gather}
    A_{X\mu'} = \Re\left[\frac{E_1}{k v}
    \frac{\begin{pmatrix}
        0& i&-\frac{\kappa_X}{\gamma } &0
    \end{pmatrix}}{\sqrt{1+\kappa_X^2/\gamma^2}}e^{k\kappa_X x'+ ik \gamma  (y'+vt')}\right].
\end{gather}
Here, we normalized the vector potential to be of amplitude $E_1/kv$. This normalization is Lorentz invariant, as a boost does not change the norm $A_\mu g^{\mu \nu} A^*_\nu$, with $A^*_\nu$ denoting complex conjugate.

Transforming back to the lab frame, and reinstating $E_1= E_1(z)$,
\begin{gather}
    A_{X\mu} =\Re\left[\frac{E_1}{kv}\frac{\begin{pmatrix}
        \kappa_X\beta & i& - \kappa_X&0
    \end{pmatrix}}{\sqrt{1+\kappa_X^2/\gamma^2}}e^{k\kappa_X x+ iky}\right].
\end{gather}

The static perturbation in the lab frame is
\begin{multline}
	    \mathbf{E}_X= \frac{E_1\kappa_X}{\sqrt{1+\kappa_X^2/\gamma^2}}e^{k\kappa_X x}\left(\kappa_X \cos ky \mathbf{e}_x-\sin ky \mathbf{e}_y\right)\\
+\pdv{E_1}{z}\frac{\kappa_X}{k\sqrt{1+\kappa_X^2/\gamma^2}}e^{k\kappa_X x}\cos ky  \mathbf{e}_z,\label{eq:E_X simplified}
\end{multline}
\begin{multline}
	    \mathbf{B}_X= B_0\frac{E_1e^{k\kappa_X x}}{E_0\sqrt{1+\kappa_X^2/\gamma^2}}(\kappa_X^2-1)\cos ky\mathbf{e}_z\\
-\frac{B_0}{E_0}\pdv{E_1}{z}\frac{e^{k\kappa_X x}}{k\sqrt{1+\kappa_X^2/\gamma^2}}\left(\kappa_X\cos ky \mathbf{e}_x-\sin ky\mathbf{e}_y\right).\label{eq:B_X simplified}
\end{multline}
The polarization of these fields is can be read of the first lines of equations (\ref{eq:E_X simplified}) and (\ref{eq:B_X simplified}). The second lines are required in order to maintain the fields as derivatives of the potentials.

The lab-frame perturbation is a pure electric multipole when $\kappa_X=1$. The perturbation shifts continuously into a magnetostatic ripple in $B_z$ as $\kappa_X \rightarrow 0$. This can be thought of as a change in the polarization of the 4-potential. This is a compressional Alfv\'en mode, occurring at $M_A=1$. 

\subsubsection{General X waves}\label{sssec: general X}

The low-frequency approximation is limited, in that it cannot take into account the details of $S$ and $D$, and as such frequency dependence,  in the dispersion relation, and as we discuss in Section \ref{sec:4}, the ponderomotive effect is strongest near the cyclotron resonances. As a result, we would like to look at the wave dynamics when the $S$ and $D$ terms of the dispersion relation vary strongly near resonance. 


When the wave is evanescent, $p$ is purely imaginary and the right and left circular polarizations suffice, but when the wave is propagating, $p$ is complex with both a real and an imaginary parts. This indicates a phase shift different than $\pi/2$ between the components. We can decompose the wave into the same right and left polarization, but remain with some (large) linearly polarized component in the $x'$ direction.

\begin{figure}
	\centering
	\includegraphics[width = \columnwidth]{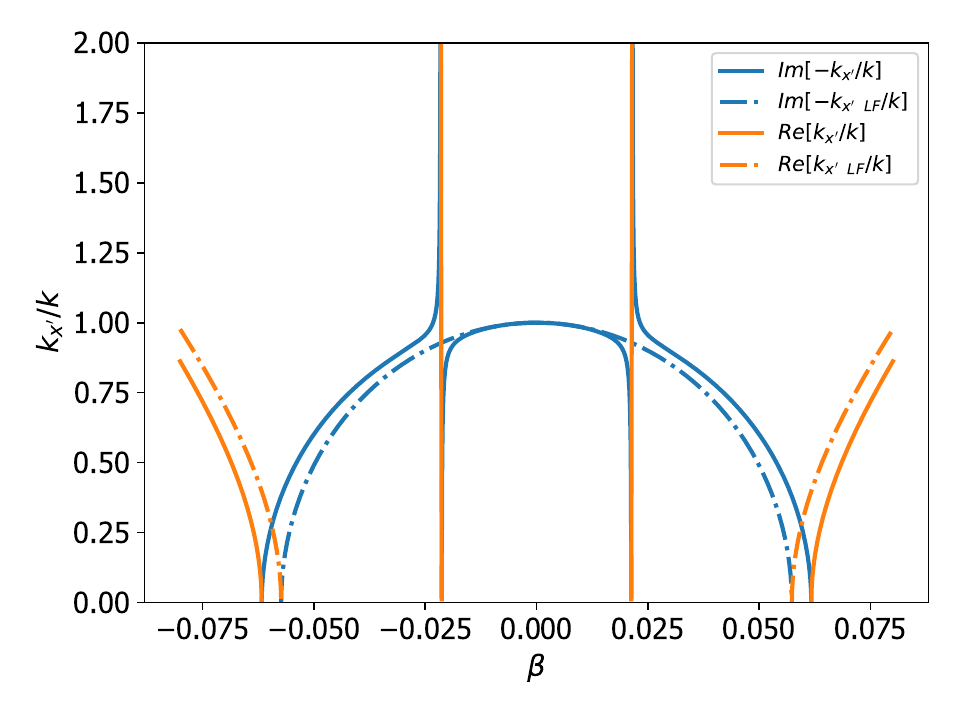}
	\caption{The exact dispersion relation, $k_{x'}$ as a function of $\beta = v/c$, for a quasi-neutral electron-proton-boron11 plasma, with $n_p = 0.5\cdot 10^{20}[m^{-3}]$, $n_{b11} = 0.1\cdot 10^{20}[m^{-3}]$, and $B_0 = 10[T]$, and $k = 100[m^{-1}]$, plotted in solid lines. The wave is evanescent in the regions plotted in blue, and is propagating in the regions plotted in orange. In here, a pair of a new cutoff and a resonance appear around $\beta = \pm 0.0215$. The low-frequency approximation is plotted in dashed lines. }\label{fig:k100epb11 plasma}
\end{figure}

\begin{figure}
	\centering
	\includegraphics[width = \columnwidth]{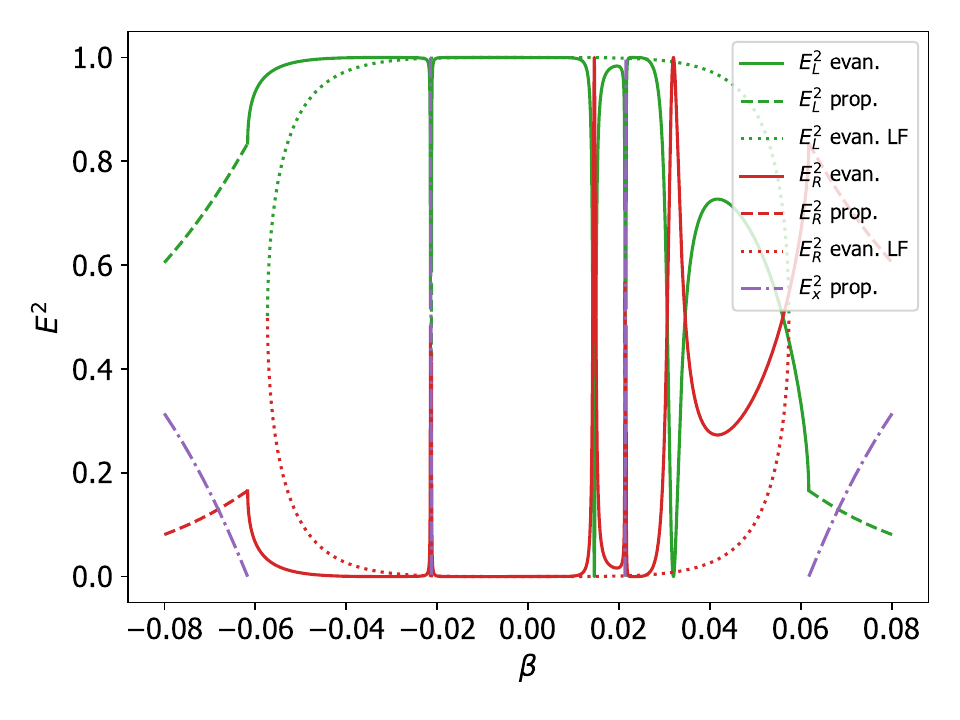}
\caption{Wave polarization squared in the moving frame, as a function of $\beta$, for the same parameters defined in Figure \ref{fig:k100epb11 plasma}. Plotted are the squares of the coefficients of the unit vector in the electric field direction. In the evanescent regime, the wave is composed of only right and left polarization, without a phase shift. In the propagating regime, the wave acquires a phase-shift, essentially splitting into three components. See Figure \ref{fig:k100RenormE1} for the ponderomotive potentials.} \label{fig:k100epb11 plasma pol}
\end{figure}

In Figure \ref{fig:k100epb11 plasma}, we plot the dispersion relation for an electron-proton-boron11 quasineutral plasma, with different curves representing the exact expressions for the evanescent and propagating regimes from equations (\ref{eq:kappa X}) and (\ref{eq:tilde k X}), as well as the approximation in equation (\ref{eq:kappa X LF}). The wave in the moving frame is in the low frequency limit when $\beta\ll 0.014$ for the parameters chosen for this plot. Here, in contrast to the electron-proton plasma, the expression for $k_{x'}$ doesn't remain accurate past the limit $|\omega'|\approx \omega_{cB11}$. In Figure \ref{fig:k100epb11 plasma pol} we plot the squares of the coefficients of the wave, for the left, right and linear polarizations. In this figure, we see that the polarization oscillates wildly at positive $\beta$s, as the wave is significantly modified by the boron11 and proton cyclotron resonances. The low-frequency approximation is not very helpful in this case. 


In Appendix \ref{app:fields X}, we write the expressions for the electric and magnetic fields in the lab frame for an X wave-like static perturbation for both the evanescent and propagating regimes.
The boundary conditions generating this perturbation are discussed in Appendix~\ref{app:2}.

\section{Ponderomotive potential}\label{sec:4}

Because practical devices would have a flow velocity of up to $\pm 0.1 c$, we shall consider non-relativistic Hamiltonian in this section.

In this section, we derive the ponderomotive potentials in dimensionless form. We measure length in units of $k^{-1}$, time in units of $\omega_{c0}^{-1}$, and momentum in units of $m\omega_{c0} /k$, for the leading order Hamiltonian, and writing the O wave perturbation cyclotron frequency $\omega_{c1}$ in units of $\omega_{c0}$, and the X wave perturbation amplitude $E_1$ in units of $B_0 \omega_{c0}/k$. This last choice, of writing
\begin{gather}
	\tilde{E}_1 =\frac{E_1k}{B_0\omega_{c0}},
\end{gather}
while having the dimensional magnetic perturbation e.g. equations (\ref{eq:B_X simplified}), (\ref{eq:B_X evan}), and (\ref{eq:B_X prop}) depending on the ratio $E_1/E_0=-\tilde{E}_1/\tilde{v}$ generate the resonance at $v=0$. An alternative choice, of writing the perturbation electric field $E_1$ in units of $E_0$ would remove this resonance at the cost of zeroing the perturbation at zero flow velocity.

The dimensionless Hamiltonian for motion in electromagnetic fields derives from the electric potential $\phi$ and vector potential $\mathbf{a}$ is
\begin{gather}
	h  =\frac{1}{2}(\tilde{\mathbf{p}}-\mathbf{a})^2+\phi.
\end{gather}

Using the generating function
\begin{gather}
	F = \frac{1}{2}\tilde{p}_x^2\cot \theta+\tilde{y}\left(\tilde{X}+\tilde{v}\right)-\tilde{p}_x \tilde{X},
\end{gather}
to change variables from the Cartesian coordinates and their conjugate momenta into the action-angle variables $\tilde{\mu},\ \theta,\ \tilde{X},\ \tilde{Y}$, with the following relations
\begin{gather}
	\tilde{x} = \sqrt{2\tilde{\mu}}\cos \theta+\tilde{X},\\
	\tilde{p}_x = -\sqrt{2\tilde{\mu}}\sin\theta,\\
	\tilde{y} = \tilde{Y}-\sqrt{2\tilde{\mu}}\sin\theta,\\
	\tilde{p}_y = \tilde{X}+\tilde{v},
\end{gather}
with $\tilde{\mu}$ being the magnetic moment or first adiabatic invariant, which is related to the gyro radius by the above expressions, $\theta$ being the gyro angle, $\tilde{Y}$ being the gyro center position, and $\tilde{X}$ its conjugate momentum. The $\tilde{z},\ \tilde{p}_z$ pair is unaffected by this transformation. In order for the particle to remain in the plasma region, the constraint on $\tilde{X}$ is $\tilde{X}<-\sqrt{2\tilde{\mu}}$.

Using these coordinates, the component $h_0$ of the Hamiltonian is solvable, with $\tilde{\mu}, \tilde{X}$ constants, and can be written down as 
\begin{gather}
	h_0 = \frac{\tilde{p}_z^2}{2}+\tilde{\mu} +\tilde{X}\tilde{v}+ \frac{1}{2}\tilde{v}^2 ,
\end{gather}
which generates gyration in the $x-y$ plane, drift in the $y$ direction, and a ballistic motion in the $z$ direction.

{Writing $\tilde{\rho} = \sqrt{2\tilde{\mu}}$ and dropping the tildes (on all quantities) for brevity.}

In order to get ponderomotive potentials, we require the perturbations to be small, i.e. $\omega_{c1}<1$ or $E_1<1$ for the two perturbations considered here. Additionally, we require a separation of scales between the $z$ ramp-up of the perturbation and the $y$ rate of interaction with the perturbation, i.e. 
\begin{gather}
	\left|\pdv{h_1}{z}v_z\right|\ll \left|\pdv{h_1}{Y}v\right|.
\end{gather}

Another requirement is for the beat period between the gyro motion harmonics and the interaction with the perturbations to be smooth
\begin{gather}
	\forall m: \frac{v_z}{m-v}\frac{1}{L} \ll1,
\end{gather}
with $L$ being the ramp-up length scale of the perturbation.

\subsection{Ponderomotive Potential for O Wave-like Perturbation}
The magnetostatic contributions to the Hamiltonian $h_{1O} = -p_z a_{1z}$, $h_{2O} = a_{1z}^2/2$ can be written in action-angle variables as
\begin{gather}
	h_{1O} = p_z\omega_{c1} e^{\kappa_O \left( \rho \cos \theta+X\right)}\cos \left(Y-\rho\sin\theta\right),\\
	h_{2O}= \frac{1}{4}\omega_{c1}^2e^{2\kappa_O  \left( \rho\cos \theta+X\right)}\left(1+\cos\left(2Y- 2\rho\sin\theta\right)\right).
\end{gather}
This is a similar expression found by\cite{karneyStochasticIonHeating1978, karneyStochasticIonHeating1979} for wave-particle interaction, with the notable difference; The wave we consider here is electromagnetic (defined by an $A$), rather than electrostatic wave (defined by a $\Phi$), generating the two terms $-e p_z A_z/m$ and $e^2 A_z^2/2m$, the second of which generates the leading order positive-definite ponderomotive potential.

We can expand the perturbation in a Fourier series
\begin{gather}\nonumber
	h_{1O} = p_z\omega_{c1} e^{\kappa_O X}\sum_{n=-\infty}^{\infty}\sum_{m=-\infty}^{\infty}I_n\left(\kappa_O\rho\right)\\\times J_{m-n}\left(\rho\right)\cos\left(m\theta - Y\right)\\
\nonumber	h_{2O}= \frac{1}{4}\omega_{c1}^2e^{2\kappa_O X}\sum_{n=-\infty}^{\infty}I_n\left(2\kappa_O\rho\right)\\\times \left(\cos n \theta+\sum_{m=-\infty}^{\infty}J_{m-n}\left(2\rho\right)\cos\left(m\theta -2 Y\right)\right).
\end{gather}
With $J_m$ being the Bessel function of the first kind of order $m$, and $I_n$ being the modified Bessel funciton of the first kind of order $n$.


Looking at the two components of the Hamiltonian, away from resonance and assuming a separation of timescales, the ponderomotive effect due to $h_{1O}$ is a modification to the effective mass for the $z$ degree of freedom at $\mathcal{O}(\omega_{c1}^2)$, and the leading order ponderomotive potential being the average over the $Y,\theta$ angles
\begin{gather}
	\phi_\mathrm{pond,O} = \frac{1}{4}\omega_{c1}^2e^{2\kappa_O X}I_0\left(2\kappa_O\rho\right),\label{eq:Phi0O}
\end{gather}
similarly to the case explored in the cylinder. 

In the limit of small gyro radius, this leading order expression, equation (\ref{eq:Phi0O}) agrees exactly with the expression found in \citet{dodinApproximateIntegralsRadiofrequencydriven2005}, using $E_\parallel = \gamma E_0 B_1e^{k \kappa_O X} / B_0$ from equation (\ref{eq:E1O}), and $I_0\left(0\right)=1$.

Seeing that the ponderomotive pseudopotential in equation (\ref{eq:Phi0O}) depends on only a single coordinate, $z$, the action of this pseudopotential would be to repel particles along $z$.

\subsection{Ponderomotive Potential for X Wave-like Perturbation}

Similarly, the contributions  $h_{1X} = -(p_y-a_{0y})a_{1y}-p_x a_{1x}+\phi$ and $h_{2X}= \mathbf{a}_1^2/2$ to the Hamiltonian for an evanecent X wave
\begin{gather}
	h=h_0+h_{1X\mathrm{evan}}+h_{2X\mathrm{evan}},
\end{gather}
can be written down in action-angle form as
\begin{multline}
h_{1X\mathrm{evan}} = -\frac{\rho E_1e^{\kappa_X( X+\rho \cos \theta)}}{2v\sqrt{|p|^2+1}}\\ 
\times \Bigg(\left(1-\frac{\Im[p]}{\gamma}\right)\cos ( \rho \sin\theta-Y+\theta)\\+\left(1+\frac{\Im[p]}{\gamma} \right)\cos (\rho \sin\theta-Y-\theta)  \Bigg),
\end{multline}
\begin{multline}
h_{2X\mathrm{evan}} =\frac{1}{4}\frac{E_1^2}{v^2}\frac{e^{2\kappa_X ( X+\rho \cos \theta)} }{|p|^2+1}\Bigg(1+\frac{\Im[p]^2}{\gamma^2} \\+\left(1 -\frac{\Im[p]^2}{\gamma^2} \right)\cos (2\rho \sin \theta-2Y)\Bigg).
\end{multline}


Expanding $h_{1X\mathrm{evan}}$ in a Fourier series
\begin{gather}
	h_{1X\mathrm{evan}} = \sum_{m=-\infty}^{\infty}\mathcal{V}_m(\mu, X)\cos\left(m\theta - Y\right),
\end{gather}
\begin{multline}
\mathcal{V}_m=-\frac{1}{2}\frac{ E_1}{v\sqrt{|p|^2+1}} \rho e^{\kappa_X  X} \sum_{n=-\infty}^{\infty} I_n\left(\kappa _X \rho\right) \times \\\left(\left(1-\frac{\Im[p]}{\gamma}\right) J_{m-n-1}\left(\rho\right)+\left(1+\frac{\Im[p]}{\gamma}\right) J_{m-n+1}\left(\rho\right)\right).
\end{multline}

And $h_{2X\mathrm{evan}}$ is
\begin{multline}
	h_{2X\mathrm{evan}} =\frac{1}{4}\frac{E_1^2}{v^2}\frac{e^{2\kappa_X X} }{|p|^2+1}\sum_{n=-\infty}^{\infty} I_n\left(2\kappa_X\rho\right)\\
\times\Bigg(\left(1 +\frac{\Im[p]^2}{\gamma^2}\right)\cos n\theta\\\left.+\left(1 -\frac{\Im[p]^2}{\gamma^2} \right)\sum_{m=-\infty}^{\infty}J_{m-n}\left(2\rho\right)\cos \left(m\theta-2Y\right)\right).
\end{multline}

We can use the perturbation method to solve for the correction to the motion due to $h_{1X}$. Using the procedure outlined in\cite{caryLieTransformPerturbation1981, depritCanonicalTransformationsDepending1969} to solve for the generating functions $w_{1m}$ such that
\begin{gather}
	\{w_{1m},h_0\}=\pdv{w_{1m}}{\theta}+v \pdv{w_{1m}}{Y} = -\mathcal{V}_m \cos\left(m\theta - Y\right),\\
	w_{1m} =-\frac{\mathcal{V}_m}{m-v} \sin \left(m\theta - Y\right),
\end{gather}

and the leading order correction to an average Hamiltonian is
\begin{multline}
	\langle h_{1X\mathrm{evan}}\rangle =\frac{1}{2}\langle\{w_{1m},h_{1\ell}\}\rangle \\= -\frac{1}{4} \frac{1}{m-v}\left(m \pdv{\mathcal{V}_m^2}{\mu}-\pdv{\mathcal{V}_m^2}{X}\right).
\end{multline}

The total leading order ponderomotive potential is
\begin{gather}\nonumber
\phi_\mathrm{pond, X, evan}=\sum_{m=-\infty}^{\infty} -\frac{1}{4} \frac{1}{m-v}\left(m \pdv{\mathcal{V}_m^2}{\mu}-2\kappa_X\mathcal{V}_m^2\right)\\+\frac{1}{4}\frac{E_1^2}{v^2}e^{2\kappa_X X}\frac{1 +\Im[p]^2/\gamma^2}{|p|^2+1} I_0\left(2\kappa_X\rho\right)\label{eq:pondx}
\end{gather}

\subsubsection{Cold particles}
For cold particles with $\rho\approx 0$, the coefficients $\mathcal{V}_1$ and $\mathcal{V}_{-1}$ contribute the most, with
\begin{gather}
\mathcal{V}_1 \approx -\frac{1}{2}\frac{1-\Im[p]/\gamma}{\sqrt{|p|^2+1}} \frac{E_1}{v} \rho e^{\kappa_X  X},\\
\mathcal{V}_{-1}  \approx -\frac{1}{2}\frac{1+\Im[p]/\gamma}{\sqrt{|p|^2+1}}\frac{E_1}{v}  \rho e^{\kappa_X X},
\end{gather}
and the rest being negligible. The expression in equation (\ref{eq:pondx}) becomes
\begin{multline}
\phi_\mathrm{pond,X}=- \frac{1}{8}\frac{E_1^2}{v^2} \frac{e^{2\kappa_X  X}}{|p|^2+1}\\\times\left[ \frac{(1-\Im[p]/\gamma)^2}{1-v}+ \frac{(1+\Im[p]/\gamma)^2}{1+v}\right]\\+\frac{1}{4}\frac{E_1^2}{v^2}  e^{2\kappa_X X} \frac{1 +\Im[p]^2/\gamma^2}{|p|^2+1}\\
=\frac{ E_1^2 e^{2\kappa_X X}}{8v(|p|^2+1)}\left[\frac{(1+\Im[p]/\gamma)^2}{1+v}- \frac{(1-\Im[p]/\gamma)^2}{1-v}\right].\label{eq:rho 0 pot}
\end{multline}
This expression agrees with the expression for the ponderomotive potential which can be found in many sources, such as \citet{dodinApproximateIntegralsRadiofrequencydriven2005}.

Returning to dimensional units, at this limit ($\rho\approx0$) and with $\gamma=1$,
\begin{multline}
\Phi_\mathrm{pond,X}=\frac{e^2B_0}{4mk E_0}\frac{ E_1^2 e^{2k\kappa_X X}}{(|p|^2+1)}\\ \times\Bigg[\left(\frac{1+\Im[p]}{\sqrt{2}}\right)^2\frac{1}{E_0k / B_0-\omega_c}\\+ \left(\frac{1-\Im[p]}{\sqrt{2}}\right)^2\frac{1}{E_0k / B_0+\omega_c}\Bigg].
\end{multline}
I.e., using their notation, the electric field can be decomposed into 
\begin{gather}
	E_+ =  E_1 e^{k\kappa_X X}\frac{1-\Im[p]}{\sqrt{2}\sqrt{|p|^2+1}},\\
	E_- =  E_1 e^{k\kappa_X X}\frac{1+\Im[p]}{\sqrt{2}\sqrt{|p|^2+1}},\\
	\omega_\mathrm{wave} = \frac{kE_0}{B_0}.
\end{gather}

{
\subsection{Effect of perturbation on the flow velocity}
Inspecting the leading order approximate Hamiltonian for either of the two perturbations
\begin{gather}
	h\approx \frac{1}{2}\left(p_z^2+v^2\right) + X v+\mu +\phi_\mathrm{pond},
\end{gather}
we can derive the gyro center flow velocity for particles interacting with the perturbation from $\dot Y = \partial h / \partial X$,
\begin{gather}
	\dot{Y} \approx  v+ 2 \kappa \phi_\mathrm{pond} .
\end{gather}
That is, particles interacting with the perturbations are pushed to larger flow velocities by the O wave analog, and to either larger or smaller velocities by the X wave analog, depending on the sign of $v$, that is flow velocity over and particle charge. This result is not restricted to cold particles only.
}


\section{Numerical validation} \label{sec:5}

In order to numerically demonstrate the validity of the ponderomotive potential expressed in equation (\ref{eq:pondx}), the Lorentz force was numerically integrated using a second-order particle pusher, which is volume preserving and generalizes Boris' method \cite{zenitaniBorisSolverParticleincell2018,boris1970relativistic,qinWhyBorisAlgorithm2013}, using the LOOPP code which was used in several of our previous publications \citep{ochsNonresonantDiffusionAlpha2021a,ochsPonderomotiveRecoilElectromagnetic2023,rubinGuidingCentreMotion2023a, rubinMagnetostaticPonderomotivePotential2023a}.

The electric and magnetic field used in the simulation are the dimensionless fields used in Appendix \ref{app:fields X}, i.e. for the evanescent regime
\begin{gather}
	\mathbf{E} = -v\mathbf{e}_x+\frac{E_1}{\sqrt{\left|p\right|^2+1}} f(z)e^{\kappa_X x}( \kappa_X\sin y\mathbf{e}_x+\cos y\mathbf{e}_y)\nonumber\\
 +\frac{E_1}{\sqrt{\left|p\right|^2+1}}\dv{f(z)}{z}e^{\kappa_X x}\sin y \mathbf{e}_z,
\\
	\mathbf{B} =\left(1-\frac{E_1}{v}f(z)\frac{ \kappa_X+\Im[p]/\gamma}{\sqrt{\left|p\right|^2+1}}e^{\kappa_X x}\sin y\right)\mathbf{e}_{z}\nonumber\\
+\frac{E_1e^{\kappa_Xx}}{v\sqrt{\left|p\right|^2+1}}\dv{f(z)}{z}\left(\sin y \mathbf{e}_x-\frac{\Im[p]}{\gamma}\cos y\mathbf{e}_y\right),
\end{gather}
with the function 
\begin{gather}
	f(z) = \begin{cases}
	0& z\le-L/2\\
	0.5 + z/L & -L/2<z<L/2\\
	1 & z\geq L/2.
\end{cases}
\end{gather}

The parameters used are $E_1 = 0.05$, in order to eliminate higher order effects, and avoid the stochastic threshold, and $L=1000$, in order to resolve the ponderomotive potentials close to resonance. The initial velocity $v_{z0} = 1$.

The simulation was terminated when the particle reached  $z>L/2$  or if $x>0$. The ponderomotive potential reported from the numerical simulation is $\phi_\mathrm{numerical}=(v_{z0}^2-v_{zf}^2)/(2 E_1^2)$ with $v_{zf}$ begin value of the $z$-velocity at the end of the simulation.

\begin{figure}
	\centering
	\includegraphics[width = \columnwidth]{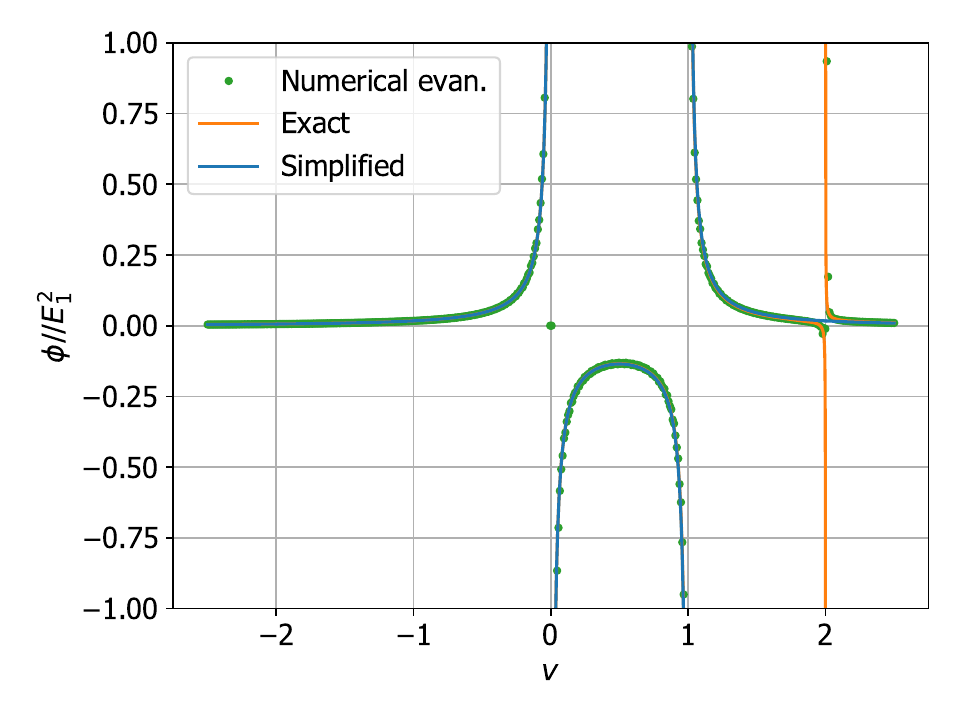}
	\caption{Dimensionless ponderomotive potentials for a vacuum field with a left polarization, at $X=-1$ and with a Larmor radius $\rho = 0.14$. The simplified potentials are nearly identical to the exact ones, which take finite gyro radius into account, except near the $v=2$ resonance. In vacuum, the perturbation is evanescent.}\label{fig:vacuumpond1}
	\includegraphics[width = \columnwidth]{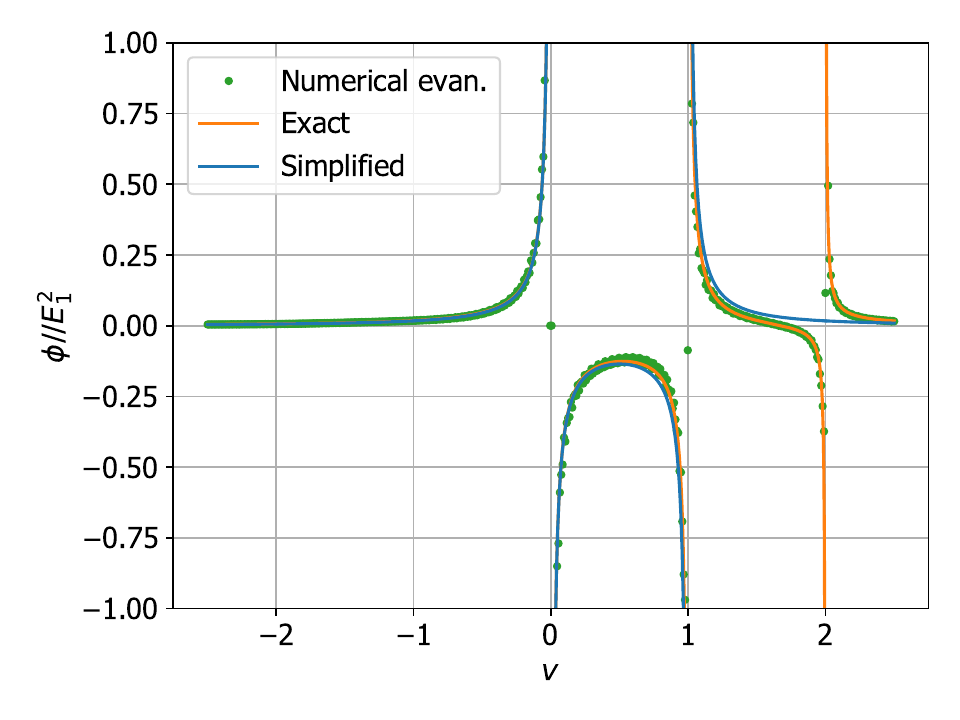}
	\caption{Dimensionless ponderomotive potentials for a vacuum field with a left polarization, at $X=-1$ and with a Larmor radius $\rho = 0.44$. This is the approximate dimensionless gyro radius for a proton with a $100[keV]$ of energy in the gyro motion degree of freedom, immersed in a $B_0=10[T]$ magnetic field and interacting with a perturbation with $k=100[m^{-1}]$. The $v=2$ resonance generates an appreciable ponderomotive potential for this gyro radius. In vacuum, the perturbation is evanescent.}\label{fig:vacuumpond2}
\end{figure}

In Figures \ref{fig:vacuumpond1} and \ref{fig:vacuumpond2} we show the agreement between the numerical simulation for vacuum fields and the expression in equation (\ref{eq:pondx}), where the sum over the indices $m,n$ was truncated to the range $-4$ to $4$ for both, for selected values of $\rho>0$. The expression for the $\rho=0$ ponderomotive potentials, equation (\ref{eq:rho 0 pot}) is plotted in blue for reference.

It is clear that even in the vacuum limit, any flow in the positive $y$ direction would generate some negative (attractive) ponderomotive potential into the perturbation region, and the potential increases as the velocity approaches the first ion cyclotron resonance. 

\begin{figure}
	\centering
	\includegraphics[width = \columnwidth]{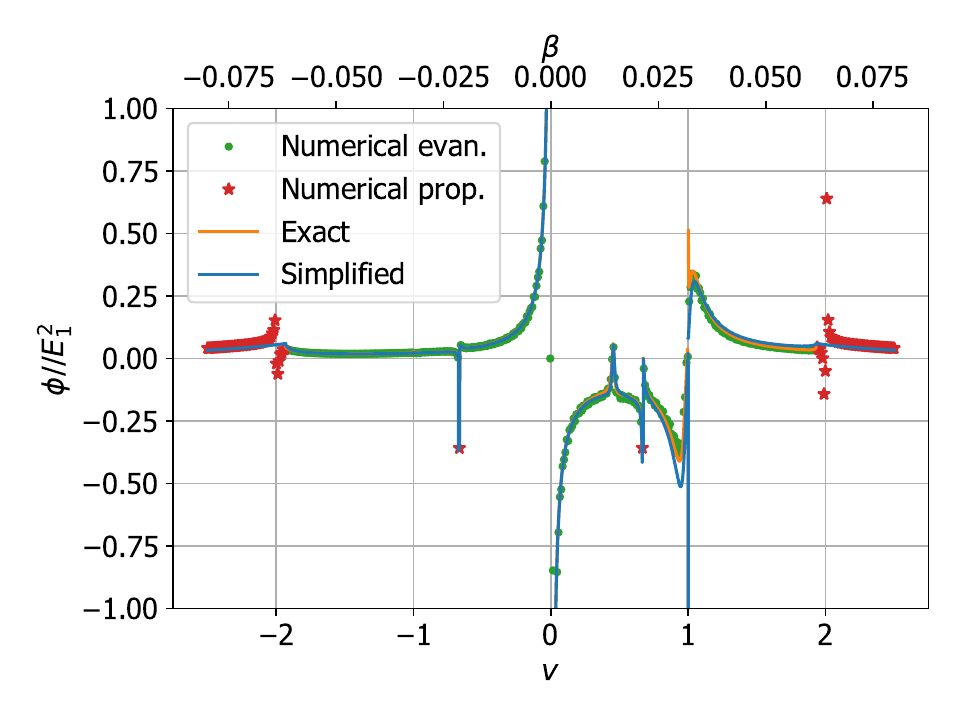}
	\caption{Numerical evaluation of the ponderomotive potential (sans the $E_1^2$ amplitude) for protons in a proton-boron11 plasma, with $n_p = 0.5\cdot10^{20}[m^{-3}]$ and $n_{B11}=0.1\cdot10^{20}[m^{-3}]$, $k=100[m^{-1}]$, $B_0=10[T]$ and $\rho= 0.3$. Notice we see here the 2nd resonances at $v=1,\pm 2$, and no resonance at $v=-1$. The green markers indicate a regime where the perturbation is evanescent, and the red markers indicate a propagating perturbation. Most of the ponderomotive potential profile can be explained by the variations in the wave polarization, which is represented in the simplified expression.}\label{fig:k100RenormE1}
\end{figure}

In Figure \ref{fig:k100RenormE1}, we see the resultant ponderomotive potential generated for the same electron-proton-boron11 plasma as in Figures \ref{fig:k100epb11 plasma} and \ref{fig:k100epb11 plasma pol} with a $k=100[m^{-1}]$, for dimensionless velocities corresponding to $-2.5$ to $2.5$ times the cyclotron frequency. The relativistic $\beta = v/c$ is plotted on top, to compare with the non-dimensional with respect to the wave, velocity v. In orange, the propagating regime has a significant contribution of a linearly polarized component, generating a resonance at the $v =\pm2$ marks. In the evanescent regime, the wave is mostly left polarized, and as such exhibits a resonance at $v=1$, but not for $v=-1$. The thin regions in which the wave propagates near the boron11 cyclotron resonance are also visible and produce a clear ponderomotive reaction from the particles.



\begin{figure}
	\centering
	\includegraphics[width = \columnwidth]{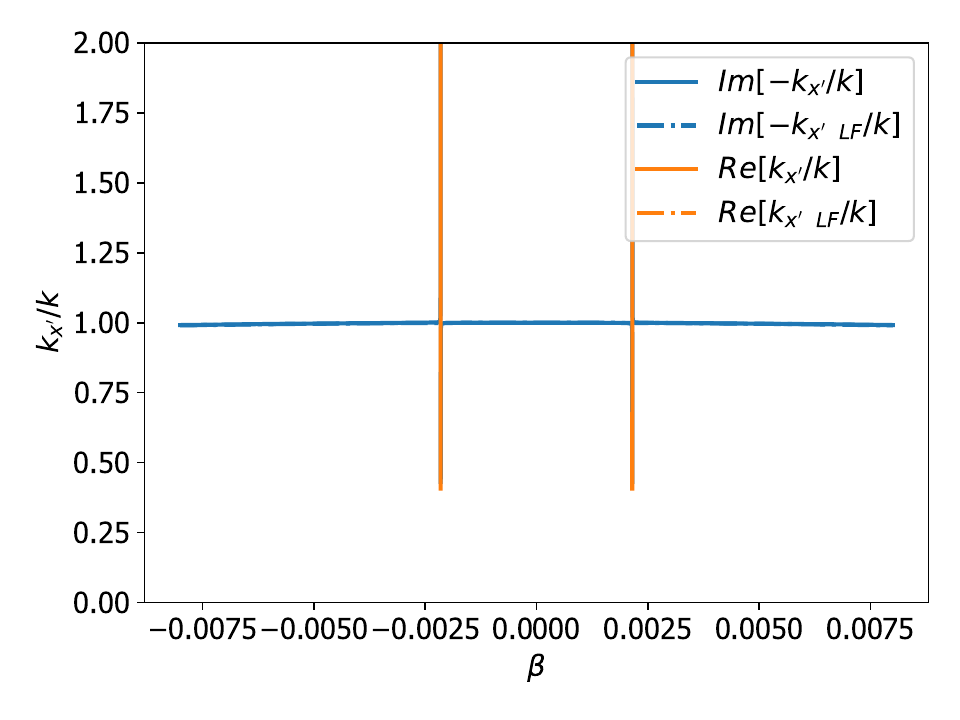}
	\caption{The exact dispersion relation, $k_{x'}$ as a function of $\beta = v/c$, for a quasi-neutral electron-proton-boron11 plasma, with $n_p = 0.5\cdot 10^{20}[m^{-3}]$, $n_{b11} = 0.1\cdot 10^{20}[m^{-3}]$, $B_0 = 10[T]$, and $k = 1000[m^{-1}]$, plotted in solid lines. The wave is evanescent in the regions plotted in blue, and is propagating in the regions plotted in orange. In here, a new cutoff appears in $\beta = \pm 0.04$, and a resonance in $\beta = \pm 0.043$. The low-frequency approximation is plotted in dashed lines. }\label{fig:k1000epb11 plasma}
\end{figure}

\begin{figure}
	\centering
	\includegraphics[width = \columnwidth]{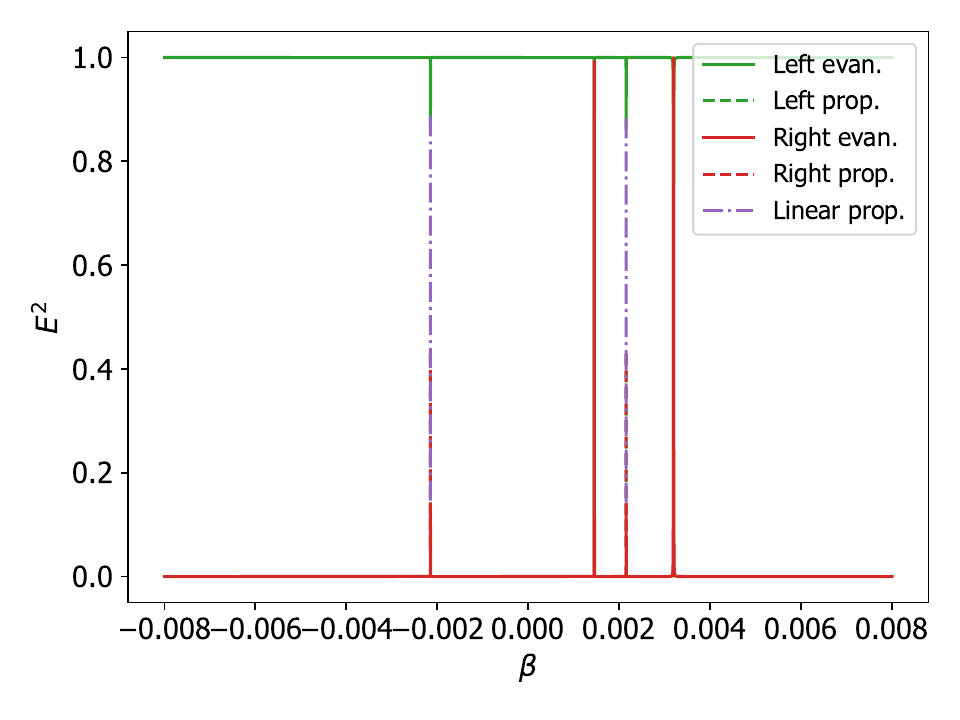}
\caption{Wave polarization squared in the moving frame, as a function of $\beta$, for the same parameters defined in Figure \ref{fig:k1000epb11 plasma}. Plotted are the squares of the coefficients of the unit vector in the electric field direction. In the evanescent regime, the wave is composed of only right and left polarization, without a phase shift. In the propagating regime, the wave acquires a phase-shift, essentially splitting into three components. See Figure \ref{fig:k1000RenormE1} for the ponderomotive potentials.} \label{fig:k1000epb11 plasma pol}
\end{figure}

\begin{figure}
	\centering
	\includegraphics[width = \columnwidth]{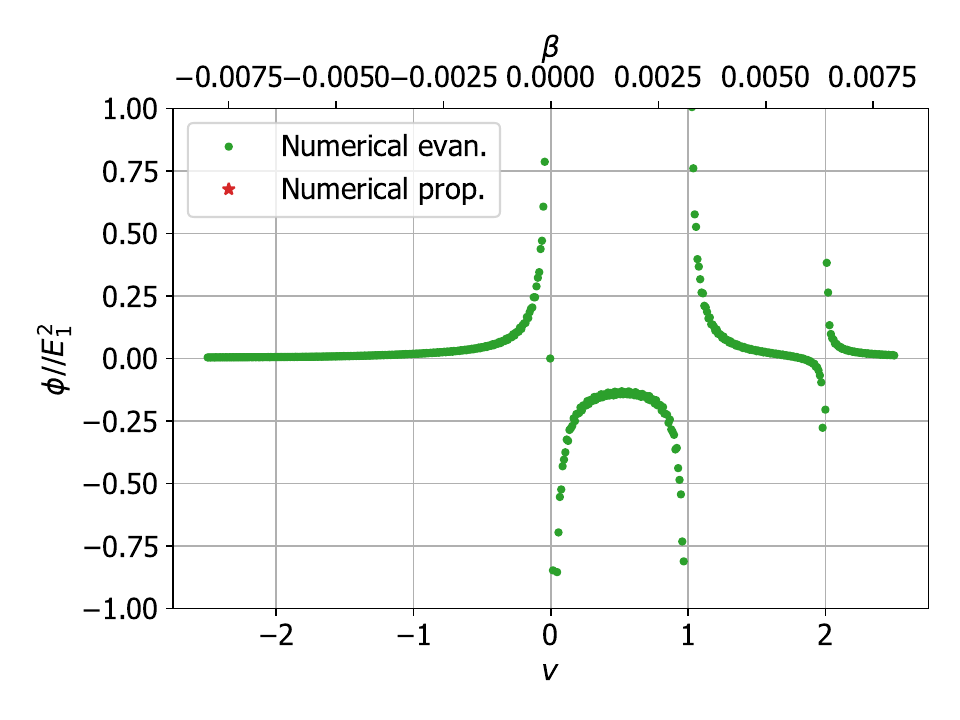}
	\caption{Numerical evaluation of the ponderomotive potential (sans the $E_1^2$ amplitude) for protons in a proton-boron11 plasma, with $n_p = 0.5\cdot10^{20}[m^{-3}]$ and $n_{B11}=0.1\cdot10^{20}[m^{-3}]$, $k=1000[m^{-1}]$, $B_0=10[T]$ and $\rho= 0.3$. Notice we see here the 2nd resonance at $v=2$, and no resonances at $v=-1,-2$.}\label{fig:k1000RenormE1}
\end{figure}

Figures \ref{fig:k1000epb11 plasma} and \ref{fig:k1000epb11 plasma pol} present the dispersion and polarization for a perturbation with a large $k=1000[m^{-1}]$, such that the flow velocity can remain small in absolute terms (i.e. compared to $c$), while interacting with the cyclotron resonances. It is evident that not much is happening at this range of $\beta$s, and the particle interaction with the wave, in Figure \ref{fig:k1000RenormE1} shows the expected resonances of a left-polarized wave. 


\section{Conclusion}

When investigating the ponderomotive potentials that can be generated by a plasma flowing past a static perturbation, two velocity scales appear in the system. The first - $\omega_c/k$ which is the cyclotron frequency of the species in question divided by the perturbation wave vector - determines the (non-relativistic) ponderomotive potential. The second is the speed of light, which together with the plasma parameters determines the plasma rearrangement due to interaction with the perturbation. 

We have used a fluid plasma model, which is a mean-field, collisionless, which is applicable for ``small signal" perturbations, and small larmor radius (cold) plasma, to investigate the plasma response. In the flute-like limit ($k_\parallel\approx 0$), these modes are equivalent to the O and X waves in the frame moving with the flow. For plasmas satisfying the above assumptions, these are the only available modes, and only a linear combination of electromagnetic field configurations are possible. 

The O wave-like perturbation is a magnetostatic perturbation with a multipolar-like $\mathbf{B}_1\perp\mathbf{B}_0$ perturbation, and is rejected from the plasma on the electron skin depth scale. This makes using the always-repulsive ponderomotive potential generated by the flow quite difficult for fusion plasmas of considerable density. The X wave-like perturbation in the vacuum limit is a multipolar-like $\mathbf{E}_1\perp\mathbf{B}_0$ perturbation, which propagates more easily in denser plasmas, and transitions smoothly into a magnetostatic perturbation $\mathbf{B}_1\parallel \mathbf{B}_0$ as the plasma becomes denser. The ponderomotive potential generated by interaction with this perturbation is a Miller-type potential which depend on the perturbation amplitude squared, with the usual resonant denominators. As such, it can have either a positive or a negative sign, depending on the flow velocity compared to the cyclotron frequency. 

We have found that the correct rotating-cylinder analog can have (mostly) a left-polarized X wave-like perturbation interacting with the plasma if it is applied via a boundary condition on the outer edge of the cylinder, and the opposite, (mostly) right-polarized X wave-like perturbation can be launched from an inner annual liner in the positive ``radial" direction. This is important if we intend to use a positively-charged plasma in an annular magnetic mirror-machine, and attempt to confine ions in a negative (i.e. attractive) potential well. 

The ponderomotive potential requires requires a separation of time-scales between the ramp-up of the perturbation and the rate of interaction with it, from the perspective of the particle. This requires a large flow velocity, large k, and a large ramp-up length scale $L_z$. At the same time, large flow velocity is detrimental for fusion applications, as it both requires energy investment to bring particles to a large energy and generates a large electric field $E_0 = v B_0$ near material boundaries. Large electric fields can damage the first walls of the vacuum vessel. However, a large k is difficult to achieve beyond around $k\approx 100[m^{-1}]$ for technological reasons, and large $k$ in the evanescent regime means the perturbation decays in a short distance from the boundary. 

A design of a ponderomotive well would have to consider the trade offs between these parameters. 

\section*{Acknowledgments}
The authors would like to thank Elijah Kolmes, and Alex Glasser for useful discussions. This work was supported by ARPA-E Grant No. DE-AR0001554 and NNSA Grant No. DE-SC0021248. This work was also supported by the DOE Fusion Energy Sciences Postdoctoral Research program, administered by the Oak Ridge Institute for Science and Education (ORISE) and managed by Oak Ridge Associated Universities (ORAU) under DOE Contract No.
DE-SC0014664.

\appendix
\section{Topological considerations for linearly polarized magnetic field}\label{app:1}

The magnetic perturbation which was investigated in \citet{ochsCriticalRoleIsopotential2023b} is \begin{gather}
	\mathbf{B}_1 = B_1(z) \cos ky \mathbf{e}_x,\label{eq:ian}
\end{gather}
where the leading order fields are as presented in Section \ref{sec:2}, $\mathbf{B}_0 = B_0 \mathbf{e}_z$ and $\mathbf{E}_\mathrm{un} = E_0  \mathbf{e}_x$. In order to get a ponderomotive potential it is required that $\partial \ln  B_1/\partial z \ll k$. For a constant $x, z$ coordinates in the lab frame, the magnitude of the perturbation changes as a function of the $y$ coordinate, but its polarization remains along the $x$ axis. In the vacuum perturbation discussed in Refs.~\onlinecite{rubinGuidingCentreMotion2023a, rubinMagnetostaticPonderomotivePotential2023a}, and its slab analog equation (\ref{eq:b1va}), the magnitude of the perturbation remains constant for a constant $x, z$ coordinates, while the polarization rotates in the $x-y$ plane. In the non-vacuum case, the perturbation has an elliptical rather than circular polarization, as discussed in Section \ref{sec:2}. 

The 4-potential describing these field in equation (\ref{eq:ian}) in the moving frame is
\begin{gather}
	A_{\mu'} = \begin{pmatrix}
		0 & 0 & 0 & \frac{B_1}{k}\sin k\gamma (y'+vt')
	\end{pmatrix},
\end{gather}
and the electromagnetic fields in this frame are
\begin{gather}
	\mathbf{E}_1' =  B_1 \gamma \frac{E_0}{B_0}\cos k\gamma (y'+vt')\mathbf{e}_{z'},\\
	\mathbf{B}_1' =  B_1\gamma \cos k\gamma (y'+vt')\mathbf{e}_{x'}.
\end{gather}
In the moving frame, this is clearly a wave propagating (primarily) perpendicularly to the magnetic field $\mathbf{B}_0' = B_0\mathbf{e}_{z'}/\gamma$. It has a wave vector component $k_{y'}=\gamma k$, frequency $\omega = -\gamma k v$ and it is purely polarized in the $z'$ direction. For this polarization, the dispersion relation is that of an O wave, meaning 
\begin{gather}
	N_{x'}^2 = P-\beta^{-2},
\end{gather}
exactly as described in Section \ref{sec:2}. Neither in vacuum nor in a fluid plasma, a linearly polarized magnetic perturbation, i.e. equation (\ref{eq:ian}), can exist when $k_z\approx 0$. This requires $\kappa_O=0$, but the expression under the square root is positive definite. 

In his book, \citet{stixWavesPlasmas1992} presents a CMA diagram for a two component plasma in Figure 2-1, with indications of the topology for the wave normal surfaces. The case discussed here is one where the wave frequency is much smaller than the ion cyclotron frequency, meaning above the $L=\infty$ line, but with a negative $\sum_s \omega_{ps}^2/\omega^2$, meaning to the left of region 12, as labeled in the diagram.

{In contrast, as found in Section \ref{sec:2}, a linearly polarized magnetic perturbation parallel to $\mathbf{B}_0$ is entirely possible, albeit for a restricted set of parameters for which $\kappa_X=0$. This is because the parallel magnetic perturbation corresponds to a different branch of the dispersion relation, which correspond to compressional Alfv\'en wave in the MHD limit.}

\section{The elecric and magnetic fields in the lab frame for an X wave-like static perturbation}\label{app:fields X}

A 4-potential generating the perturbation in the lab frame is
\begin{gather}
    A_{\mu}  =\Re\left[ \frac{E_1}{k v\gamma }\frac{i\begin{pmatrix}
		- \beta \gamma &p & \gamma  &0
	\end{pmatrix}}{\sqrt{\left|p\right|^2+1}}e^{ik_{x'}x+ ik y}\right],
\end{gather}
with $p$ being the wave polarization in the moving frame, as used in Section \ref{sssec: general X}.

Taking $E_1\in \mathbb{R}$, the static perturbation itself for the evanescent regime (with $p = i(\beta^2 \gamma D+\kappa_x)/(\beta^2 \gamma S-\gamma)$) is
\begin{multline}
	\mathbf{E}_{X\mathrm{evan}}= \frac{E_1 e^{k\kappa_X x}}{\sqrt{\left|p\right|^2+1}} \left( \kappa_X\sin ky\mathbf{e}_x+\cos ky\mathbf{e}_y\right)\\
+\frac{e^{k\kappa_X x}}{k\sqrt{\left|p\right|^2+1}}\pdv{E_{1}}{z}\sin ky\mathbf{e}_z,
\end{multline}
\begin{multline}
	\mathbf{B}_{X\mathrm{evan}}= B_0\frac{E_1}{E_0}\frac{ \kappa_X+\Im[p]/\gamma}{\sqrt{\left|p\right|^2+1}}e^{k\kappa_X x}\sin ky\mathbf{e}_{z}\\
+\frac{B_0e^{k\kappa_Xx}}{kE_0\sqrt{\left|p\right|^2+1}}\pdv{E_1}{z}\left(-\sin ky \mathbf{e}_x+\frac{\Im[p]}{\gamma}\cos ky\mathbf{e}_y\right).
\end{multline}
 
For the propagating regime (with $p = (i\beta^2 \gamma D-\tilde{k}_x)/(\beta^2 \gamma S-\gamma)$), the static perturbation is
\begin{multline}
	\mathbf{E}_{X\mathrm{prop}}= \frac{E_1}{\sqrt{|p|^2+1}} \left( \tilde{k}_X\mathbf{e}_x+\mathbf{e}_y\right)\cos k (\tilde{k}_X x+y)\\
+\frac{1}{k\sqrt{|p|^2+1}}\pdv{E_{1}}{z}\sin k (\tilde{k}_X x+y) \mathbf{e}_z,
\end{multline}
\begin{multline}
	\mathbf{B}_{X\mathrm{prop}}= B_0\frac{E_1}{E_0}\frac{\tilde{k}_X-\Re[p]/\gamma}{\sqrt{|p|^2+1}}\cos k (\tilde{k}_X x+y) \mathbf{e}_z\\
 -B_0\frac{E_1}{E_0}\frac{1}{\sqrt{|p|^2+1}}\frac{\Im[p]}{\gamma }\sin k (\tilde{k}_X x+y) \mathbf{e}_z\\
+\frac{B_0}{kE_0\sqrt{|p|^2+1}}\pdv{E_1}{z}\left(\frac{\Re[p]}{\gamma}\mathbf{e}_y- \mathbf{e}_x\right)\sin k (\tilde{k}_X x+y) \\
+\frac{B_0}{kE_0\sqrt{|p|^2+1}}\pdv{E_1}{z}\frac{\Im[p]}{\gamma }\cos k (\tilde{k}_X x+y)\mathbf{e}_y.\label{eq:B_X prop}
\end{multline}

\section{Boundary conditions for the X wave-like perturbation}\label{app:2}

In this appendix we look at the X wave on the $x>0$ side of the boundary, for a plasma with the same $\mathbf{E}_0$ and $\mathbf{B}_0$ and but allowing for a different composition. This plasma is the slab-analogue to a plasma rotating outside a cylinder. 

For an evanescent $N_{x'}^2<0$ X wave on the $x>0$ side of the boundary, the wave vector component has the opposite sign as in the $x<0$ side of the boundary,
\begin{gather}
	k_{x'+} =ik \kappa_{X+}.
\end{gather}
The wave polarization is similarly modified,
\begin{gather}
	p_+=\frac{h_{x'+}}{h_{y'+}} = i\frac{D-\frac{\kappa_{X+}}{\gamma\beta^2}}{S-\frac{1}{\beta^2}},\ \ h_{z'+}  =0. 
\end{gather}

The fields for $x>0$, for the evanescent regime are
\begin{multline}
	\mathbf{E}_{X\mathrm{evan}+}= \frac{E_{1+} e^{-k\kappa_{X+} x}}{\sqrt{\left|p_+\right|^2+1}} \left( -\kappa_{X+}\sin ky\mathbf{e}_x+\cos ky\mathbf{e}_y\right)\\
+\frac{e^{-k\kappa_{X+} x}}{k\sqrt{\left|p_+\right|^2+1}}\pdv{E_{1+}}{z}\sin ky\mathbf{e}_z,
\end{multline}
\begin{multline}
	\mathbf{B}_{X\mathrm{evan}+}= B_0\frac{E_{1+}}{E_0}\frac{- \kappa_{X+}+\Im[p_+]/\gamma}{\sqrt{\left|p_+\right|^2+1}}e^{-k\kappa_{X+} x}\sin ky\mathbf{e}_{z}\\
+\frac{B_0e^{-k\kappa_{X+}x}}{kE_0\sqrt{\left|p_+\right|^2+1}}\pdv{E_{1+}}{z}\left(-\sin ky \mathbf{e}_x+\frac{\Im[p_+]}{\gamma}\cos ky\mathbf{e}_y\right).\label{eq:B_X evan}
\end{multline}

If the perturbation at the two sides of the boundary $x=0$ is evanescent, the relation between the fields is as follows. Across the $x=0$ boundary, the electric field perpendicular to the boundary and the magnetic field normal to the boundary are continuous, yielding the relation between the amplitudes,
\begin{gather}
E_{1-} =E_{1+} \sqrt{\frac{\left|p_-\right|^2+1}{\left|p_+\right|^2+1}}.
\end{gather}
The surface charge distribution on the boundary is related to the jump in the electric field perpendicular to the boundary
\begin{gather}
\sigma = -\epsilon_0E_{1-} \frac{\kappa_{X+} + \kappa_{X-}}{\sqrt{\left|p_-\right|^2+1}}\sin ky.
\end{gather}
The surface currents on the $x=0$ boundary that generate the jump in the magnetic field is
\begin{multline}
	\mathbf{J}  =\delta(x)\frac{B_0}{\mu_0E_0}\frac{E_{1-}}{\sqrt{\left|p\right|^2+1}}\\\times\left( \kappa_{X+}+	\kappa_{X-}+\frac{\Im[p_-]-\Im[p_+]}{\gamma}\right)\sin ky \mathbf{e}_y.
\end{multline}
This current has a non-zero divergence and requires another component to make it divergence-less, for example with a component going in the positive or negative x direction.

For a propagating X wave on the $x>0$ side, the wave vector similarly switches sign. 
\begin{gather}
	k_{x'+} =-k \tilde{k}_X.
\end{gather}
The polarization of a propagating X wave is 
\begin{gather}
	p_+=\frac{h_{x'+}}{h_{y'+}} = \frac{iD+ \frac{\tilde{k}_X }{\gamma \beta^2}}{S-\frac{1}{\beta^2}},\ \ h_{z'+}  =0. 
\end{gather}

\section*{References}
%

\end{document}